\documentclass{iau} 
\usepackage{graphicx}
\newcommand{\object}[1]{}
\newcommand{\farcm}{\mbox{\ensuremath{.\mkern-4mu^\prime}}}

\title[Runaway O stars in Gaia DR1] %% give here short title %%
{New runaway O-type stars in the \\ first Gaia Data Release}

\author[Jes\'us Ma{\'\i}z Apell\'aniz et al.]   %% give here short author list %%
{Jes\'us Ma{\'\i}z Apell\'aniz$^1$,
      Rodolfo H. Barb\'a$^2$,
      Sergio Sim\'on-D{\'\i}az$^{3,4}$,
      Ignacio Negueruela$^5$,
 \and Emilio Trigueros P\'aez$^{1,6}$}

\affiliation{$^1$Centro de Astrobiolog{\'\i}a, CSIC-INTA, Spain \\
             email: {\tt jmaiz@cab.inta-csic.es} \\[\affilskip]
             $^2$Universidad de La Serena, Chile \\[\affilskip]
             $^3$Instituto de Astrof{\'\i}sica de Canarias, Spain \\[\affilskip]
             $^4$Universidad de La Laguna, Spain \\[\affilskip]
             $^5$Universidad de Alicante, Spain \\[\affilskip]
             $^6$Universidad Complutense de Madrid, Spain}

\pubyear{2017}
\volume{329}  %% insert here IAU Symposium No.
\jname{The lives and death-throes of massive stars}
\editors{A.C. Editor, B.D. Editor \& C.E. Editor, eds.}
\begin{document}

\maketitle

\begin{abstract}
We have detected 13 new runaway-star candidates of spectral type O combining the TGAS
(Tycho-Gaia Astrometric Solution) proper motions from Gaia Data Release 1 (DR1) and the
sample from GOSSS (Galactic O-Star Spectroscopic Survey). We have also combined TGAS and
Hipparcos proper motions to check that our technique recovers many of the previously
known O-type runaways in the sample. 
%% add here a maximum of 10 keywords, to be taken form the file <Keywords.txt>
\end{abstract}

\object{AE Aur}
\object{HD 155913}
\object{zeta Pup}
\object{ALS 18929}
\object{zeta Oph}
\object{HD 104565}
\object{HD 57682}
\object{ALS 11244}
\object{mu Col}
\object{HD 155775}
\object{HD 157857}
\object{HDE 229232}
\object{Y Cyg}
\object{HD 46573}
\object{HD 116852}
\object{BD +60 134}
\object{V479 Sct}
\object{HD 12323}
\object{HD 124979}
\object{CPD -34 2135}
\object{HD 36879}
\object{HD 94024}
\object{68 Cyg}
\object{HD 192639}
\object{HD 17520}
\object{AB Cru}
\object{BD -14 5040}
\object{HD 41997}
\object{HD 152623}
\object{HD 201345}
\object{lambda Cep}
\object{alpha Cam}
\object{HD 192281}
\object{HD 175876}
\object{HD 75222}
\object{HD 14633}
\object{HD 153919}
\object{HD 195592}
\object{HD 96917}
\object{BD -08 4617}
\object{9 Sge}
\object{xi Per}
\object{HD 93521}
\object{HD 108}
\object{HDE 227018}

\firstsection % if your document starts with a section,
              % remove some space above using this command.
\section{Introduction}

On 14 September 2016 the first Gaia Data Release (DR1) was presented (\cite[Brown et al. 2016]{B16}). 
Gaia DR1 includes parallaxes and proper motions from TGAS (Tycho-Gaia Astrometric Solution) for the 
majority (but not all) of the Tycho-2 stars. The excluded Tycho-2 stars include all of the very bright 
objects but also some dimmer ones. TGAS proper motions exist for a significantly larger number of stars
than for Hipparcos and, for the stars in common between both catalogs, they are more precise.

The Galactic O-Star Spectroscopic Survey (GOSSS, \cite[Ma{\'{\i}}z Apell{\'a}niz 2011]{M11}) is 
obtaining $R\sim$2500, high-S/N, blue-violet spectroscopy of all optically accessible Galactic O stars.
To this date, three survey papers 
\cite[(Sota et al. 2011, 2014; Ma{\'{\i}}z Apell{\'a}niz et al. 2016)]{S11,S14,M16} have been published
with a total of 590 O stars. Several additional hundreds have already been observed and will be published 
in the near future.

\section{Data and methods}

Our initial plan with Gaia DR1 was to analyze the parallaxes in order to increase the meager number of
useful trigonometric distances available for O stars 
(\cite[van Leeuwen 2007; Ma{\'{\i}}z Apell{\'a}niz \etal\ 2008]{v07,M08}). However, the TGAS parallaxes 
for O stars provide little new information, as the brightest O stars are not included and only one star,
AE~Aur, has $\pi_{\rm o}/\sigma_\pi >$ 6, where $\pi_{\rm o}$ is the observed parallax and $\sigma_\pi$
is the parallax uncertainty. It should be remembered that, in general,
$<\!d\!>\; \ne 1/\pi_{\rm o}$, that is, the inverse of the observed parallax is not an unbiased estimator of 
the trigonometric distance (\cite[Lutz \& Kelker 1973; Ma{\'{\i}}z Apell{\'a}niz 2001, 2005]{L73,M01,M05c}).  

On the other hand, the TGAS proper motions proved to have useful information. By cross-matching TGAS and GOSSS 
(including unpublished objects) we found 525 Galactic O stars with proper motions, of which we discarded 5 due 
to their large uncertainties. For the unmatched GOSSS stars we searched for Hipparcos proper motions and 
discovered another 96 objects, of which 7 were discarded for the same reason. That left us with a 
total of 520 + 89 = 609 Galactic O stars with good TGAS or Hipparcos proper motions (of those, 427 are in the 
published GOSSS papers).

The proper motions in RA ($\mu_\alpha$) and declination ($\mu_\delta$) were transformed into their 
equivalents in Galactic latitude ($\mu_b$) and longitude ($\mu_l$). A robust mean for $\mu_b$ (reflecting
the solar motion in the vertical direction), $<\!\mu_b\!>$, and a robust standard deviation,
$\sigma_{\mu_b}$, were calculated. For $\mu_l$ we robustly fitted a functional form  
$f(l) = a_0 + a_1\cos l + a_2\cos 2l$ and we also calculated the robust standard deviation, $\sigma_{\mu_l}$,
from the fit. Results for stars with good TGAS or Hipparcos proper motions are shown in Fig.~1. 

\begin{figure}[t]
\begin{center}
\includegraphics[width=1.2\linewidth]{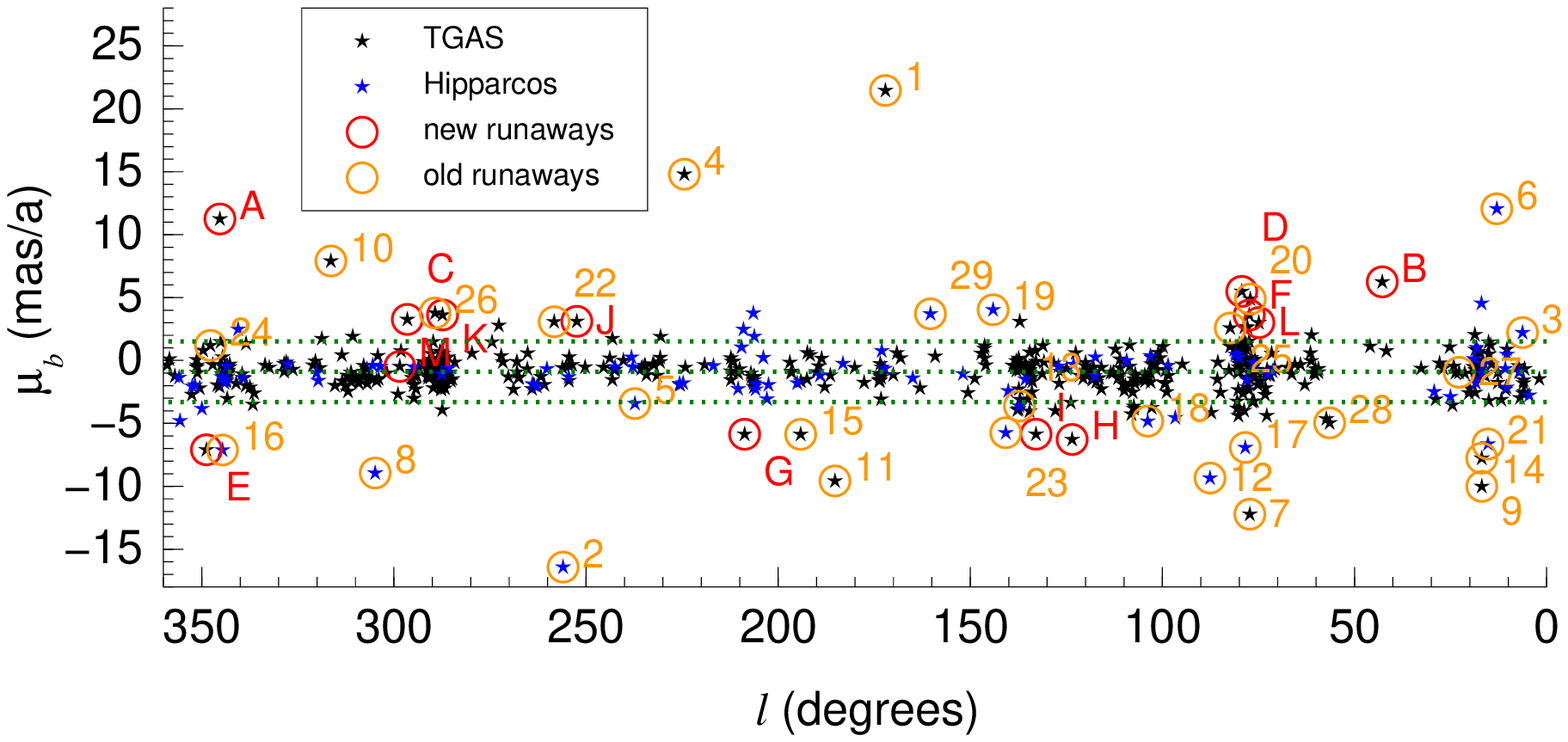}
\includegraphics[width=1.2\linewidth]{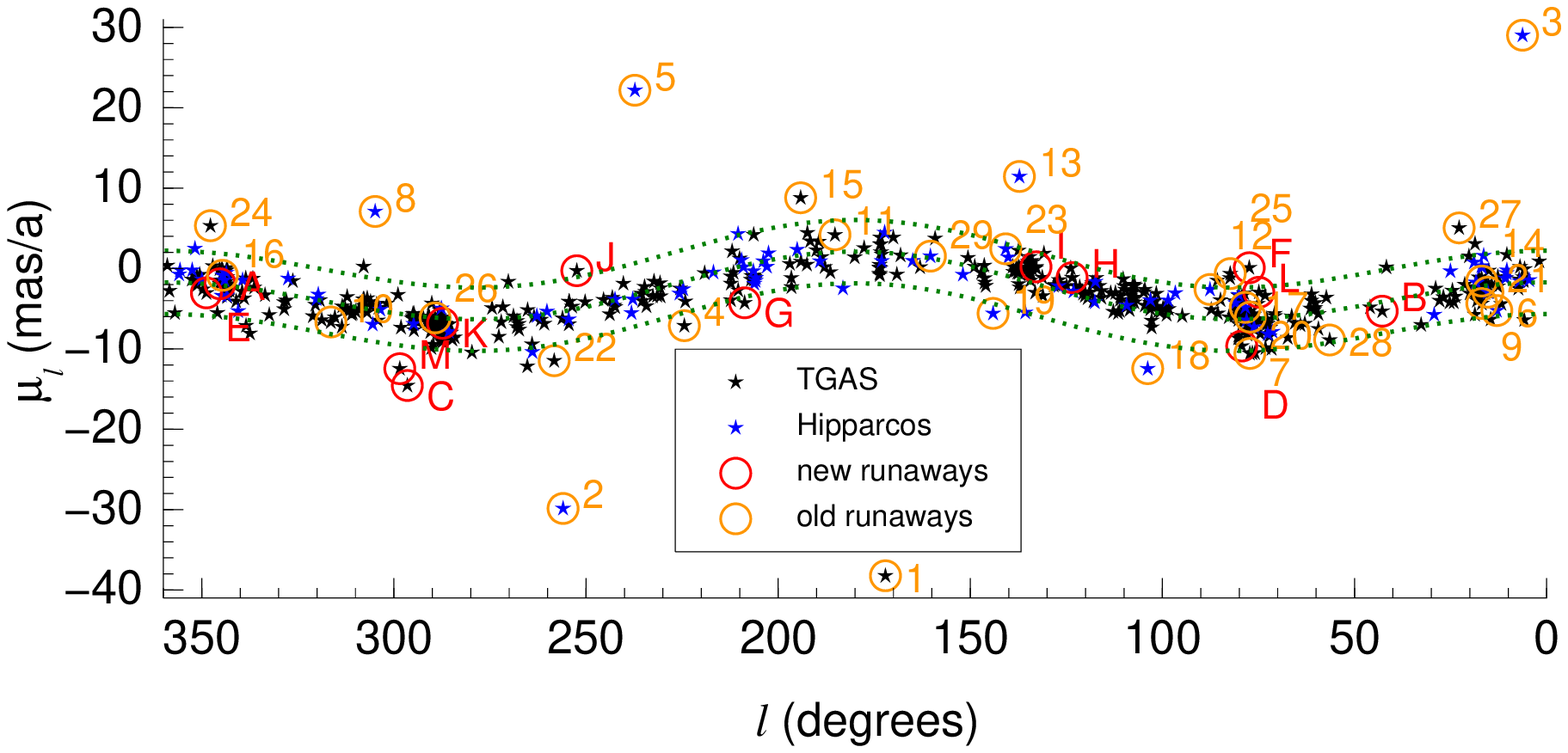}
\end{center}
\caption{Galactic latitude (top) and longitude (bottom) TGAS (black) and Hipparcos (blue) proper motions 
for the O-star sample in the sample. Circles and IDs identify runaway candidates, both new
(red, letters) and previously known (orange, numbers). See Table 1 for ID correspondences. The dotted 
green lines represent the functions and 2$\sigma$ deviations used to detect runaway candidates.}
\label{fig1}
\end{figure}

To detect runaway stars we computed the normalized difference (in standard deviations) of the difference between 
the observed proper motions and the fitted ones i.e.:

\begin{displaymath}
\Delta = \sqrt{\left(\frac{\mu_b^\prime}{\sigma_{\mu_b}}\right)^2 + \left(\frac{\mu_l^\prime}{\sigma_{\mu_l}}\right)^2},
\end{displaymath}

\noindent where $\mu_b^\prime = \mu_b-<\!\mu_b\!>$ and $\mu_l^\prime = \mu_l-f(l)$ are the corrected proper motions, 
and sorted the results from largest to smallest (Table~1). This 2-D method is simpler than a full computation 
of the 3-D velocities (e.g. \cite[Tezlaff \etal\ 2011]{T11}) and can yield false positives and
negatives (see below), but it has the advantage of being self-contained and, therefore, less prone to errors introduced by
the required external measurements in the 3-D method (distances and radial velocities).

\section{Results}

The runaway star candidates detected by our 2-D method are shown in Table~1, divided in previously known and new objects.
The cut in $\Delta$ is the same in both cases and was empirically established at 3.5 (the stars at the top of the lists, 
AE~Aur and HD~155\,913, have values of 27.6 and 10.1, respectively) by comparing our results with those of \cite{T11}. 

\begin{table}
\caption{Previously known O-type runaway stars (left) and new candidates (right). Each list is sorted
by $\Delta$, the normalized deviation from the mean latitude and longitude proper motions for their 
Galactic longitude. An ID (a number for previously known objects, a letter for new ones) is provided
for each star in order to identify it in Fig.~1. The T/H flag indicates the origin of the proper motions
(TGAS or Hipparcos, respectively). All new candidates have TGAS proper motions. Reference codes are listed
in the bibliography. GP stands for Galactic Plane. The corrected proper motions are in mas/a.}
\label{tab1}
\begin{center}
\begin{tabular}{rlcccclrrl}
\hline
\multicolumn{4}{c}{Previously known} & $\;\;\;$& \multicolumn{5}{c}{New candidates} \\
\cline{1-4} 
\cline{6-10}\\[-1.0ex]
ID & Name              & T/H & Ref. & & ID & Name            & $\mu_b^\prime$ & $\mu_l^\prime$ & Possible origin         \\
\cline{1-4} 
\cline{6-10}\\[-1.0ex]
 1 & AE Aur            & T   & H01  & & A  & HD 155\,913     &   12.14       &    0.17         & NGC 6322                \\
 2 & $\zeta$ Pup       & H   & H01  & & B  & ALS 18\,929     &    7.14       & $-$1.15         & GP, $10.6^{\rm o}$ away \\
 3 & $\zeta$ Oph       & H   & H01  & & C  & HD 104\,565     &    4.16       & $-$8.80         & GP, $4.0^{\rm o}$ away  \\
 4 & HD 57\,682        & T   & M04  & & D  & ALS 11\,244     &    6.38       & $-$3.34         & Cyg OB2                 \\
 5 & $\mu$ Col         & H   & H01  & & E  & HD 155\,775     & $-$6.20       & $-$1.10         & ---                     \\
 6 & HD 157\,857       & H   & M04  & & F  & HDE 229\,232 AB &    4.50       &    6.38         & NGC 6913                \\
 7 & Y Cyg             & T   & M05a & & G  & HD 46\,573      & $-$4.97       & $-$4.66         & GP, $2.6^{\rm o}$ away  \\
 8 & HD 116\,852       & H   & M04  & & H  & BD +60 134      & $-$5.38       &    2.04         & Cas OB7                 \\
 9 & V479 Sct          & T   & R02  & & I  & HD 12\,323      & $-$4.96       &    2.13         & Per OB1                 \\
10 & HD 124\,979       & T   & T11  & & J  & CPD -34 2135    &    4.00       &    4.74         & ---                     \\
11 & HD 36\,879        & T   & M04  & & K  & HD 94\,024      &    4.52       & $-$0.64         & Carina Nebula           \\
12 & 68 Cyg            & H   & T11  & & L  & HD 192\,639     &    3.90       &    3.28         & Dolidze 4               \\
13 & HD 17\,520 A      & H   & M04  & & M  & AB Cru          &    0.42       & $-$6.83         & ---                     \\
14 & BD -14 5040       & T   & G08  & &    &                 &               &                 &                         \\
15 & HD 41\,997        & T   & M04  & &    &                 &               &                 &                         \\
16 & HD 152\,623 AaAbB & H   & M05b & &    &                 &               &                 &                         \\
17 & HD 201\,345       & H   & T11  & &    &                 &               &                 &                         \\
18 & $\lambda$ Cep     & H   & H01  & &    &                 &               &                 &                         \\
19 & $\alpha$ Cam      & H   & M05b & &    &                 &               &                 &                         \\
20 & HD 192\,281       & T   & M04  & &    &                 &               &                 &                         \\
21 & HD 175\,876       & H   & T11  & &    &                 &               &                 &                         \\
22 & HD 75\,222        & T   & T11  & &    &                 &               &                 &                         \\
23 & HD 14\,633 AaAb   & H   & T11  & &    &                 &               &                 &                         \\
24 & HD 153\,919       & T   & M04  & &    &                 &               &                 &                         \\
25 & HD 195\,592       & T   & T11  & &    &                 &               &                 &                         \\
26 & HD 96\,917        & T   & T11  & &    &                 &               &                 &                         \\
27 & BD -08 4617       & T   & M04  & &    &                 &               &                 &                         \\
28 & 9 Sge             & T   & M04  & &    &                 &               &                 &                         \\
29 & $\xi$ Per         & H   & H01  & &    &                 &               &                 &                         \\
\hline
\end{tabular}
\end{center}
\end{table}

To check for false negatives in our list, we searched \cite{T11} for runaway candidates with 
$P_{v_{\rm pec}} >$ 0.5 missing in Table~1 but present in our sample. There are 33 objects missing but, 
of those, 30 were detected by \cite{T11} based mainly on their radial velocities as they have 
(a) $P_{v_{\rm r,pec}} > P_{v_{\rm t,pec}}$ and (b) $P_{v_{\rm t,pec}} < 0.5$. One of the remaining three 
objects is HD~93\,521, the highest - by far - latitude Galactic O star ($b = 52^{\rm o}$), which is
difficult to detect in a 2-D method designed for objects near the Galactic Plane. The other two, HD~108
and HDE~227\,018, have TGAS proper motions with significantly smaller uncertainties and closer to the
mean values than the Hipparcos values, which were the ones used by \cite{T11}. Hence, a 3-D 
reanalysis would likely reduce their $P_{v_{\rm t,pec}}$. Therefore, we conclude that our method 
correctly picks up those runaway stars with large tangential velocities but, as expected, misses some 
which are moving mostly in a radial direction.

What about false positives? The final answer will lie, of course, in future work, but there is a good reason
why the new 13 objects had not been detected before as runaways. Eight of them do not have Hipparcos proper
motions and the remaining five were not included in \cite{T11}. Another indirect evidence in favor of the reality
of the runaway condition for the new candidates is that in several cases it is possible to trace back the past 
motion of the star through its corrected proper motion to a cluster or an association as its possible origin 
(Table 1). Note that three of the new candidates are in the Cygnus region of the Galactic Plane (ALS~11\,244, 
HDE~229\,232~AB, and HD~192\,639). For HD~46\,573 we detect a bow shock in WISE images whose relative position 
with respect to the star is consistent with the corrected proper motion (Fig.~2).

\begin{figure}[t]
\begin{minipage}[t]{8.5cm}
\includegraphics[width=\linewidth]{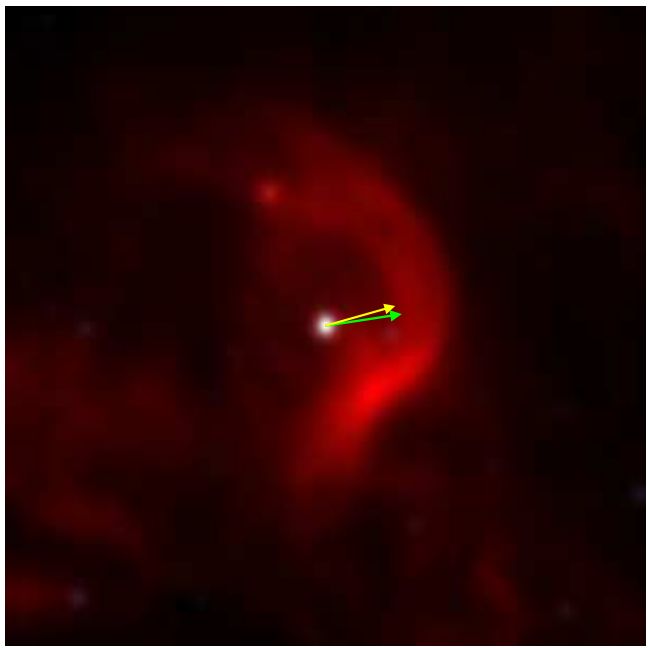}
\end{minipage} $\;\;\;\;$
\begin{minipage}[t]{4cm}
\vspace{-8.5cm}
\caption{WISE W3+W2+W1 RGB mosaic of HD~46\,573. The green arrow indicates the TGAS original proper motion and the
yellow arrow the corrrected proper motion (see text). Note how the bow shock correctly aligns with the proper motion.
The field size is $6\farcm87\times 6\farcm87$ with N to the top and E to the left.}
\end{minipage} \
\label{fig2}
\end{figure}

\section{Future work}

Our plans for the incoming years are:

\begin{itemize}
 \item Calculate extinction corrections (both $E(4405-5495)$ and $R_{5495}$) for all the stars in the sample with
       CHORIZOS (\cite[Ma{\'{\i}}z Apell{\'a}niz 2004]{M04}) using the \cite{M14} family of extinction laws in order 
       to obtain accurate spectroscopic parallaxes and compare them with the Gaia trigonometric parallaxes.
 \item Extend the analysis to the rest of the OB stars. This will allow us to search for additional runaway stars and
       redo the study of the spatial distribution of OB stars in the solar neighborhood of \cite{M01} with the much 
       better Gaia data.
 \item Use the multiepoch OWN (\cite[Barb\'a \etal\ 2010]{B10} and contribution by the same author in these 
       proceedings), IACOB (\cite[Sim\'on-D{\'\i}az \etal\ 2015]{S15} and contribution by the same author in these 
       proceedings), and CAF\'E-BEANS (\cite[Negueruela \etal\ 2015]{N15}) data to obtain radial velocities for OB stars 
       corrected for binarity as a necessary step to accurately calculate their 3-D velocity.
 \item Expand the GOSSS sample by observing new stars.
 \item Incorporate the results from the new Gaia Data Releases.
\end{itemize}

\end{document}